\documentclass[conference]{IEEEtran}

\usepackage{graphicx}
\usepackage{amsmath}
\usepackage{cite}
\usepackage{hyperref}
\usepackage{booktabs}
\usepackage{array} 
\usepackage{booktabs} 
\usepackage{makecell}
\usepackage{amsmath}
\usepackage{amsfonts} 
\usepackage{amssymb} 
\usepackage{subfig}
\usepackage{float}

\title{The Causal Impact of Dean's List Recognition on Academic Performance: Evidence from a Regression Discontinuity Design}
\author{
    \IEEEauthorblockN{Luc (Zhilu) Chen}
    \IEEEauthorblockA{
        John A. Paulson School of Engineering and Applied Sciences\\
        Harvard University\\
        Cambridge, MA, USA\\
        Email: lucchen@g.harvard.edu
    }
}

\begin{document}

\maketitle

\begin{abstract}
This study examines the causal impact of being placed on the Dean's List, a positive education incentive, on future student performance using a regression discontinuity design. The results suggest that for students with low prior academic performance and who are native English speakers, there is a positive impact of being on the Dean’s List on the probability of getting onto the Dean's List in the following year. However, being on the Dean's List does not appear to have a statistically significant effect on subsequent GPA, total credits taken, dropout rates, or the probability of graduating within four years. These findings suggest that a place on the Dean's List may not be a strong motivator for students to improve their academic performance and achieve better outcomes.
\end{abstract}

\section{Introduction and Motivation}
Having good academic performance is not only important for individual students, but also for the broader society and economy. It has been shown that higher levels of education are associated with better employment opportunities, higher wages, and improved health outcomes \cite{oreopoulos2013making}. Moreover, academic success can lead to improved critical thinking, problem-solving, and communication skills, all of which are essential in various professions \cite{veerasamy2019relationship}. Additionally, students who perform well in school are more likely to have greater self-confidence and a positive sense of self-efficacy, which can have a positive impact on their personal and professional lives \cite{pintrich2004conceptual}. Therefore, understanding the factors that contribute to academic success, such as the impact of the Dean's List, is important for educators, policymakers, and employers alike.

To enhance student academic performance and positively influence economic outcomes, educators are investigating various strategies to incentivize student achievements. One strategy is the implementation of a Dean's List, a widely adopted practice in universities that acknowledges and celebrates students' academic accomplishments. To earn a spot on the Dean's List, students are required to attain a specific grade point average (GPA) determined by the university. This recognition serves as a testament to their academic success and can inspire students to sustain or elevate their academic performance going forward.

The purpose of this study is to examine the impact of being included in the Dean's List on academic outcomes, as responses to different types of incentives can vary among individuals. The recognition can serve as a positive tool that encourages students to strive for higher achievement, which may lead to increased motivation and more effort towards completing courses and graduating on time \cite{bliven2021impact}. However, it is also important to note that such incentives may not always lead to the intended positive outcomes. For instance, research has shown that the competitiveness of certain academic programs, such as the Indiana Choice Scholarship Program, can create undue pressure on students, ultimately negatively affecting their future academic achievement \cite{canbolat2021long}. Understanding the impact of academic recognition on student motivation and achievement can have implications beyond the academic setting, including workplace motivations. Therefore, the findings of this research could inform our understanding of the factors that promote or hinder success in a variety of contexts.

This paper was inspired by Lindo et al.'s study \cite{lindo2010ability}, which used regression discontinuity design to examine the effects of negative incentives on future academic performance. In contrast, this paper utilizes regression discontinuity design (RDD) to investigate the impact of being placed on the Dean's List after the first academic year on students' continued academic performance, particularly during the freshman year. RDD is a suitable method to study this research question as it allows us to isolate the effect of being on the Dean's List from other factors that may influence academic performance. By comparing the academic performance of students just above and below the Dean's List cutoff, we can estimate the causal effect of being on the Dean's List. 

To ensure the validity of RDD by addressing the problem of nonrandom sorting, this paper conducts a balance check by investigating the distribution of student grades relative to the cutoff. The continuous distribution of grades across the cutoff indicates the successful randomization of individuals into the treatment and control groups. Moreover, we examined whether students' observable traits, such as prior academic performance, age, and gender, remained continuous through the threshold. The absence of significant discontinuity through the threshold provides strong evidence that students with particular characteristics were unable to manipulate their grades to be included in the Dean's List.

To measure academic performance, this paper uses dependent variables such as the probability of getting on the Dean’s List in the second year, sophomore year GPA, dropout rate, the number of course credits taken in the following year, and the likelihood of graduating within four years. These measures are relevant because they provide a comprehensive understanding of academic performance. The probability of getting on the Dean’s List indicates the sustainability of academic success, sophomore year GPA measures academic improvement, and the number of credits taken in the following year captures increased academic engagement and motivation. Additionally, the dropout rate reflects the likelihood of degree completion, and graduating within four years is a significant milestone that also minimizes additional tuition costs and allows earlier entry into the workforce. By considering these measures, we can gain a more comprehensive view of the impact of being on the Dean's List on students' academic outcomes.

The results of this study suggest that being on the Dean's List may not have a statistically significant effect on subsequent GPA, total credit taken, dropout rates, or the probability of graduating within four years for the overall student population. However, we did find a positive effect for students with low prior academic performance and for those who are native English speakers. For this group, being on the Dean's List increased the likelihood of sustained academic success, as they were more likely to get on the Dean's List again in the following year. These findings suggest that the Dean's List may be a more effective motivator for students who face academic challenges or do not have language barriers. However, overall, the effect of the Dean's List on academic outcomes is minor for the entire student population. These results challenge the assumption that positive incentives such as the Dean's List alone can lead to better academic performance and success. Alternative strategies may be needed to motivate students to achieve better outcomes.

\section{Literature Review}
Previous research has extensively explored the impact of negative academic incentives, such as academic probation, on student behavior and performance. Studies show that academic probation after the first semester can improve short-term academic performance; however, these effects tend to fade over time and do not significantly increase graduation rates or persistence \cite{fletcher2017effects}. In contrast, the focus of this study is on the impact of positive academic incentives.

Positive incentives have shown promise in improving academic outcomes. For instance, the El Dorado Promise scholarship, which provides financial assistance to high school graduates, has been demonstrated to have a statistically significant positive impact on math achievement \cite{ash2021promise}. Similarly, Leuven et al. found that monetary incentives offered on first-year entrance exams positively influenced students' future academic achievements, suggesting that tangible rewards can enhance motivation and performance \cite{leuven2010effect}. While the impact of monetary incentives is well-documented, research on non-monetary incentives remains limited.

One notable study by Lavy and Sand \cite{lavy2018origins} examined the effects of winning a math competition in Israel, where no financial rewards were given. The findings indicated an improvement in subsequent math test scores, underscoring the potential of non-monetary incentives to positively influence student motivation and academic performance.

The specific non-monetary incentive investigated in this study is placement on the Dean's List. Previous research by Seaver and Quarton \cite{seaver1976regression} showed that earning a spot on the Dean's List early in the academic year helped students maintain the quality of their academic work. However, it did not lead to an increase in the amount of work undertaken, as measured by the number of credits. Their study used regression discontinuity analysis on a sample of 1,002 students from Pennsylvania State University (196 on the Dean's List and 816 not on the Dean's List). While they found an improvement in subsequent GPA for Dean's List recipients, the small sample size and lack of controls for confounding factors such as age, gender, and prior academic performance limit the generalizability of their findings.

This study addresses these limitations by incorporating a larger sample size and more comprehensive controls, including student characteristics like age, gender, and prior academic performance. Additionally, it examines the longer-term outcomes of Dean's List placement, such as the likelihood of graduating within four years. Another novel aspect of this study is the exploration of gender differences in response to educational incentives, as prior research suggests that female students may be more responsive to positive incentives \cite{angrist2002new}.

In summary, while prior studies provide valuable insights into the role of incentives in academic settings, the literature lacks a comprehensive analysis of the long-term effects of non-monetary academic incentives like the Dean's List. This research aims to fill this gap by leveraging regression discontinuity design (RDD) to isolate the causal impact of Dean's List recognition on academic performance, while addressing key limitations of earlier studies.

\section{Data and Context}
The dataset used in this study comes from a Canadian university with three campuses: one main campus and two branch schools. The dataset was also used in a previous study titled "Ability, Gender, and Performance Standards: Evidence from Academic Probation" \cite{lindo2010ability}. The criteria to qualify for the Dean's List are the same across all three campuses: completion of a minimum of 5 total credits and a cumulative grade point average (CGPA) above 3.5 on a 0-4.3 scale. Students who earn placement on the Dean's List receive a congratulatory letter acknowledging their excellent academic performance. Importantly, the Dean's List does not offer tangible benefits such as monetary rewards or access to additional resources. As a result, any observed effects can be attributed solely to psychological factors, such as the sense of accomplishment and recognition associated with being placed on the list.

The dataset spans a nine-year period from 1996 to 2005 and contains 44,362 observations at the student level, with the academic year broken into fall, winter, and summer terms. The data includes key variables such as Year 1 GPA, Year 2 GPA, total credits completed, gender, age, first language, previous high school performance, ethnicity, and term registration status. Binary variables were constructed to indicate whether students qualified for the Dean's List in each respective year.

In this study, specific parameters were applied to ensure the reliability of results. Only students who completed at least two academic years were included. Sample restrictions recommended by \cite{lindo2010ability} were also implemented, limiting the sample to students who entered the university between ages 17 and 21, comprising 99\% of the remaining sample. To reduce bias and improve precision, students within 0.7 grade points of the Dean's List cutoff were selected, excluding those with exceptionally high or low academic performance.

\begin{figure}[ht]
    \centering
    \includegraphics[width=1\linewidth]{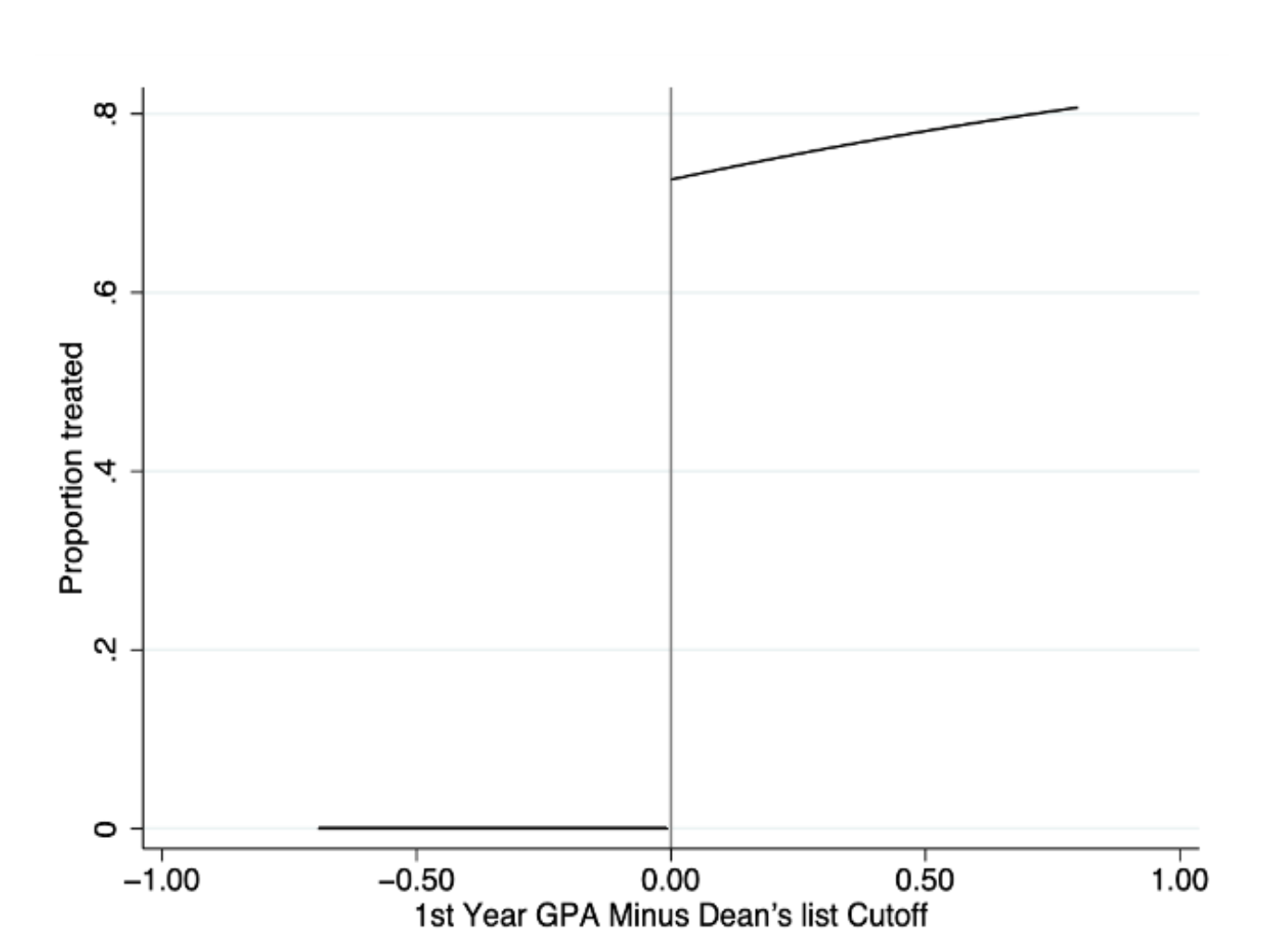}
    \caption{Predicted Probability of Getting into Dean's List}
    \label{fig:grade_distribution}
\end{figure}

Based on Figure~\ref{fig:grade_distribution}, students on the left side of the cutoff met the criteria for treatment, while only a subset on the right side did due to additional requirements, such as completing at least 5 credits. To address this non-compliance issue, students failing to meet the unit requirement after the first academic year were excluded from the analysis, ensuring comparability between treatment and control groups and reducing potential bias.

Table~\ref{tab:summary_statistics} presents summary statistics for the full sample and the restricted sample. The restricted sample includes students within 0.7 grade points of the Dean's List cutoff, who completed at least two academic years and took a minimum of five units. Compared to the full sample, the restricted sample displays notable differences in both characteristics and outcomes.

For example, the mean high school grade percentile in the restricted sample is 74.07\%, significantly higher than the full sample mean of 50.17\%. Similarly, the mean first-year GPA in the restricted sample is 3.37, compared to 2.44 for the full sample, indicating that the restricted sample comprises more academically accomplished students.

In terms of outcomes, the mean second-year GPA in the restricted sample is 3.23, compared to 2.55 for the full sample. However, the restricted sample shows a higher proportion of students leaving the university after the first year (9\% versus 5\%). Conversely, a greater proportion of students in the restricted sample graduate within four years (68\%) compared to the full sample (45\%).

By focusing on students within 0.7 grade points of the Dean's List cutoff, the restricted sample ensures greater comparability between treatment and control groups in terms of observed and unobserved characteristics. This restriction reduces potential bias and improves the precision of the treatment effect estimates, thereby increasing the validity of the causal inference from the regression discontinuity design (RDD).

\begin{table*}[h!]
\centering
\caption{Summary Statistics for the Full Sample and Restricted Sample}
\label{tab:summary_statistics}

\begin{tabular}{l>{\centering\arraybackslash}p{2.2cm}>{\centering\arraybackslash}p{2.5cm}>{\centering\arraybackslash}p{2.5cm}>{\centering\arraybackslash}p{3cm}}
\toprule
\textbf{Variables} & \multicolumn{2}{c}{\textbf{Full Sample}} & \multicolumn{2}{c}{\textbf{Restricted Sample}} \\
\cmidrule(lr){2-3} \cmidrule(lr){4-5}
& \textbf{Mean} & \textbf{SD} & \textbf{Mean} & \textbf{SD} \\
\midrule
\textit{Characteristics} \\[1ex]
High School Grade Percentile & 50.17 & 28.86 & 74.07 & 21.95 \\
First-Year GPA & 2.44 & 0.89 & 3.37 & 0.35 \\
Age at Entry & 18.67 & 0.74 & 18.55 & 0.72 \\
First-Year Credit Total & 4.57 & 0.51 & 5.01 & 0.09 \\
Male & 0.38 & 0.49 & 0.39 & 0.49 \\
Primary Language: English & 0.71 & 0.45 & 0.74 & 0.44 \\
Place of Birth: North America & 0.87 & 0.34 & 0.89 & 0.31 \\
Attend Campus 1 & 0.58 & 0.49 & 0.75 & 0.44 \\
Attend Campus 2 & 0.17 & 0.38 & 0.09 & 0.29 \\
Attend Campus 3 & 0.24 & 0.43 & 0.16 & 0.37 \\
Dean's List Eligibility & 0.09 & 0.29 & 0.38 & 0.48 \\
\\
\textit{Outcomes} \\[1ex]
Second-Year GPA & 2.55 & 0.83 & 3.23 & 0.54 \\
Left University after 1st Year & 0.05 & 0.22 & 0.09 & 0.29 \\
Second-Year Credit Total & 3.86 & 1.55 & 4.36 & 1.35 \\
Graduated by Year 4 & 0.45 & 0.50 & 0.68 & 0.47 \\
Second-Year Dean's List Eligibility & 0.09 & 0.29 & 0.26 & 0.44 \\
\bottomrule
\multicolumn{5}{p{0.85\linewidth}}{\footnotesize \textbf{Notes:} The summary statistics for the full sample and restricted sample are presented in the table. The restricted sample comprises students who meet two criteria: they are within 0.7 of the Dean's List cutoff and take more than 5 units in a single academic year. Regarding all students characteristics variables, the full sample and restricted sample contain 44362 and 10978 each. The second year GPA variable includes 38,576 students for the full sample and 9,978 for the restricted sample. Similarly, the second year credit taken variables include 43,593 in the full sample and 10,846 in the restricted sample. The graduate within four-year variable is observed for 30,017 students in the full sample and 7,167 students in the restricted sample.} \\
\end{tabular}
\end{table*}

\section{Empirical Specification and Identification}

This study aims to estimate the causal impact of being named to the Dean's List on future academic performance using a regression discontinuity design (RDD). The following equation is used for the estimation:

\begin{align}
Y_{i}^{\text{year } j} &= \beta(GPASDIZE_{i}^{\text{year } 1}) + \tau(GPASDIZE_{i}^{\text{year } 1} > 0) \notag \\
&\quad + X_{i}'\gamma + u_i
\end{align}

In this equation, \( Y_{i}^{\text{year } j} \) represents the outcome variable of interest for student \(i\) in year \(j\), such as second-year GPA, credits taken, dropout rate, probability of making the Dean’s List in the second year, or probability of graduating within four years. The variable \( GPASDIZE_{i}^{\text{year } 1} \) denotes the standardized first-year GPA for student \(i\), which measures the distance between their GPA and the Dean's List cutoff. The term \( GPASDIZE_{i}^{\text{year } 1} > 0 \) is an indicator variable that equals one if the student's GPA exceeds the cutoff and zero otherwise, capturing the treatment effect. The vector \( X_{i}' \) includes control variables such as gender, ethnicity, and prior academic performance to account for observable differences among students. The random error term \( u_i \) captures unobserved characteristics affecting \( Y_{i}^{\text{year } j} \).

The primary parameter of interest, \(\tau\), represents the causal effect of being placed on the Dean's List on the specified outcomes. A statistically significant value of \(\tau\) would indicate a discontinuity in outcomes at the cutoff that is attributable to the treatment.

The standardized distance variable, \( GPASDIZE_{i}^{\text{year } 1} \), is calculated as:

\[
GPASDIZE_{i}^{\text{year } 1} = \frac{GPA_{i}^{\text{year } 1} - \text{Cutoff}}{\sigma_{GPA}}
\]

where \( GPA_{i}^{\text{year } 1} \) is the observed first-year GPA for student \(i\), \(\text{Cutoff}\) is the GPA threshold for Dean's List eligibility (3.5 in this study), and \(\sigma_{GPA}\) is the standard deviation of first-year GPA in the sample. This calculation ensures that the variable is dimensionless, enabling comparisons across observations. A positive value of \( GPASDIZE \) indicates that a student's GPA is above the cutoff, while a negative value indicates the opposite.

To estimate the treatment effect, a local linear regression with triangular kernel weights is employed. This method prioritizes observations near the cutoff while reducing the influence of those farther away. The kernel weights are defined as:

\begin{equation}
K(x) = \max(1 - |x|, 0)
\end{equation}

where \(x = \frac{GPA_{i}^{\text{year } 1} - \text{Cutoff}}{h}\) is the normalized distance of student \(i\)'s GPA from the cutoff, and \(h\) represents the bandwidth. Observations beyond the bandwidth receive zero weight, while those closer to the cutoff are weighted more heavily, with the highest weight assigned to the cutoff itself. A bandwidth of 0.5 grade points is used in the primary analysis, and sensitivity analyses with alternative bandwidths are conducted to ensure robustness.

The weighted regression framework minimizes the following objective function:

\begin{align}
\min_{\beta, \tau} \sum_{i} 
& \; K\left(\frac{GPA_{i}^{\text{year } 1} - \text{Cutoff}}{h}\right) \notag \\
& \times \Bigg[
    Y_{i}^{\text{year } j} 
    - \beta \, GPASDIZE_{i}^{\text{year } 1} \notag \\
& \quad - \tau \, \mathbb{I}(GPASDIZE_{i}^{\text{year } 1} > 0) 
    - X_{i}'\gamma 
\Bigg]^2
\end{align}

where \(K\) is the triangular kernel weighting function, \(\mathbb{I}(GPASDIZE_{i}^{\text{year } 1} > 0)\) is the treatment indicator, and \(X_{i}'\gamma\) are the covariates. This framework ensures that observations near the cutoff, which are most relevant for causal inference, contribute more to the estimation of the treatment effect.

Given that GPA is recorded in discrete intervals (hundredths of a grade point), observations within the same GPA range may exhibit correlation. To address this, standard errors are clustered following the methodology of Lee and Card (2008). Clustering accounts for within-group correlation, ensuring reliable statistical inference.

\section{Validity of the Regression Discontinuity Design}

Before applying the Regression Discontinuity Design (RDD) method, it is essential to ensure its validity. Addressing the issue of non-random sorting is crucial, as intentional positioning above or below the cutoff could bias results. In this study, we investigate whether students may have manipulated their GPA to meet the Dean's List criteria. Several factors mitigate this concern. First, first-year students are often less familiar with university policies and may not know the GPA required to be on the Dean's List. Second, final grades are based on end-of-semester exams, making it challenging for students to adjust their performance in time to meet the threshold.

We examine the distribution of student grades relative to the threshold using a local bandwidth of 0.05 grade points. Figure~\ref{fig:grade_distribution} shows the estimated density of GPA scores, with no significant discontinuities observed. This suggests no evidence of sorting or manipulation near the cutoff, ensuring the balance between treatment and control groups.

\begin{figure}[ht]
\centering
\includegraphics[width=1\linewidth]{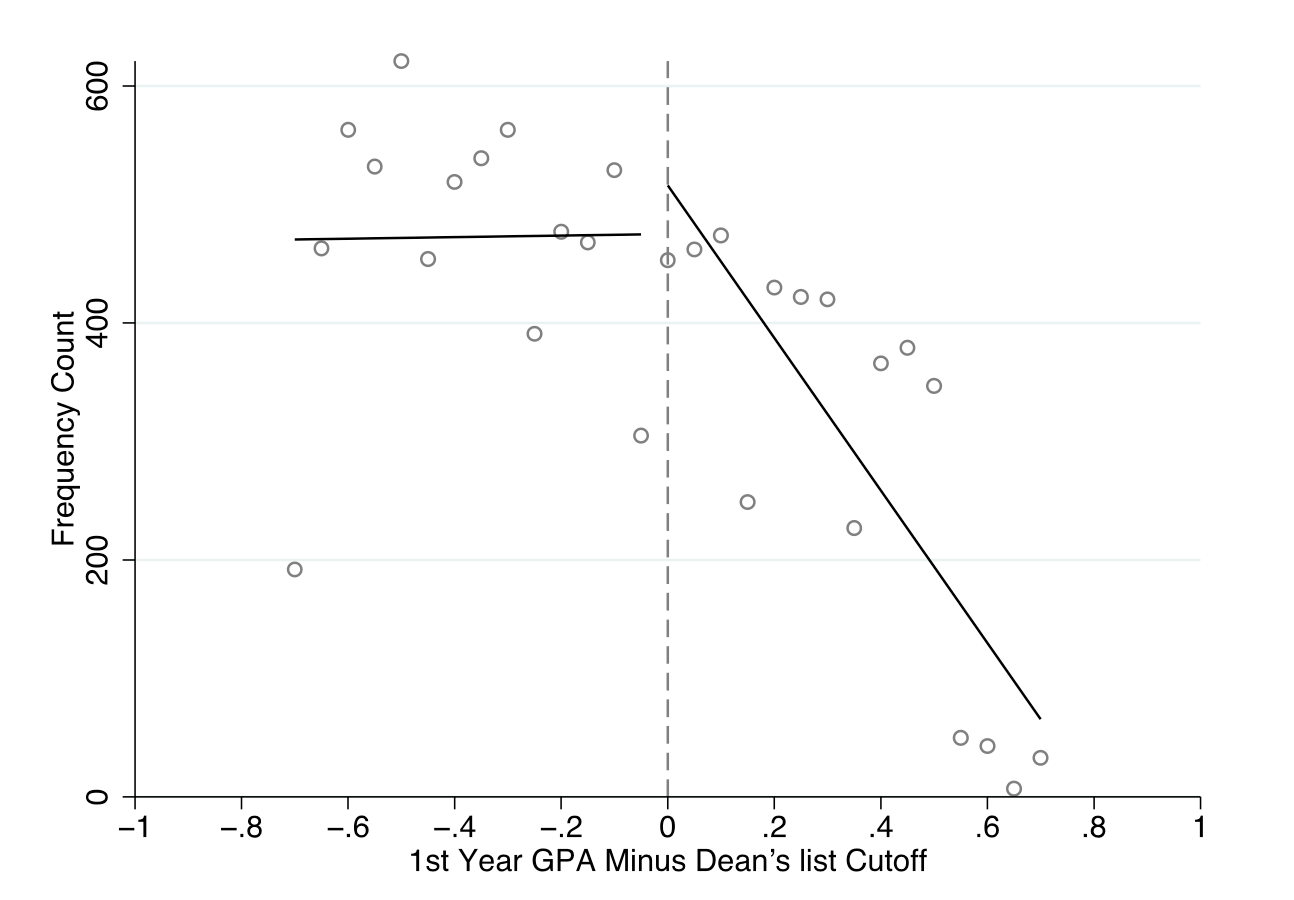} 
\caption{Student Grade Distribution Relative to the Cutoff}
\label{fig:grade_distribution}
\end{figure}

In addition to ensuring no sorting near the cutoff, we verify that observable characteristics remain continuous across the threshold. Table~\ref{tab:discontinuities} presents the results of these checks. We observe no significant differences in high school grade percentile, age at entry, gender, birthplace, language, credits taken, or campus attendance near the threshold. These results confirm that the treatment and control groups are comparable, supporting the validity of the RDD framework.

\begin{table*}[ht]
\centering
\caption{Estimated Discontinuities in Students' Observed Traits}
\label{tab:discontinuities}
\def\sym#1{\ifmmode^{#1}\else\(^{#1}\)\fi}
\begin{tabular}{l*{8}{c}}
\hline\hline
\addlinespace
                &\multicolumn{1}{c}{\makecell{HS grade \\ percentile \\ ranking}}&
                \multicolumn{1}{c}{\makecell{Age \\ at \\ entry}}&
                \multicolumn{1}{c}{\makecell{Male}}&
                \multicolumn{1}{c}{\makecell{Borned in \\ North \\ America}}&
                \multicolumn{1}{c}{\makecell{Native \\ English \\ Speaker}}&
                \multicolumn{1}{c}{\makecell{Credit \\  Taken}}&
                \multicolumn{1}{c}{\makecell{Attending \\ Campus \\ 1}}&
                \multicolumn{1}{c}{\makecell{Attending \\ Campus \\ 2}}\\ 
\midrule
First year GPA $>$ cutoff & 0.104 & 0.016 & -0.003 & -0.018 & -0.013 & -0.021 & 0.013 & 0.008 \\
                         & (1.127) & (0.031) & (0.018) & (0.013) & (0.021) & (0.068) & (0.011) & (0.071) \\
\addlinespace
Constant                & 76.7\sym{***} & 18.57\sym{***} & 0.386\sym{***} & 0.909\sym{***} & 0.75\sym{***} & 0.757\sym{***} & 0.085\sym{***} & 0.158\sym{***} \\
                         & (0.653) & (0.018) & (0.011) & (0.009) & (0.016) & (0.039) & (0.007) & (0.04) \\
\midrule
Observations            & 10978 & 10978 & 10978 & 10978 & 10978 & 10978 & 10978 & 10978 \\
\hline\hline
\multicolumn{9}{l}{\footnotesize Notes: Standard errors in parentheses, clustered on GPA.}\\
\multicolumn{9}{l}{\footnotesize \sym{*} \(p<0.10\), \sym{**} \(p<0.05\), \sym{***} \(p<0.01\).}\\
\end{tabular}
\end{table*}

\section{Results}

This study examines the impact of Dean's List placement on several key academic outcomes, including the probability of making the list in the second year, subsequent GPA, total credits taken, dropout rate, and likelihood of graduating within four years. These outcomes serve as various measures of academic success and provide a comprehensive understanding of the relationship between Dean's List recognition and overall academic achievement. Furthermore, the analysis explores differential effects across student subgroups based on high school academic performance, gender, and native language proficiency.

The findings highlight variations in the response to Dean's List recognition. Prior research suggests that students from disadvantaged backgrounds or with poor prior academic performance tend to be more responsive to educational incentives \cite{angrist2009incentives, cismaru2022impact}. Additionally, differences in gender responses to incentives are expected, as women are generally more receptive to positive reinforcements such as scholarships and academic advising \cite{angrist2009incentives}. Non-native English speakers may face additional challenges in academic environments, which could influence their response to Dean's List placement. The results of this study aim to shed light on these disparities and provide actionable insights into how institutions can support student success.

\subsection{Probability of Making the Dean's List in the Second Year}

Table~\ref{tab:table3} explores the effect of Dean's List recognition in the first year on the likelihood of being on the list again in the second year. The results indicate that for the overall student population, the impact is small and statistically insignificant. However, the subgroup analysis reveals notable variations.

\begin{figure}[h!]
\centering
\includegraphics[width=\linewidth]{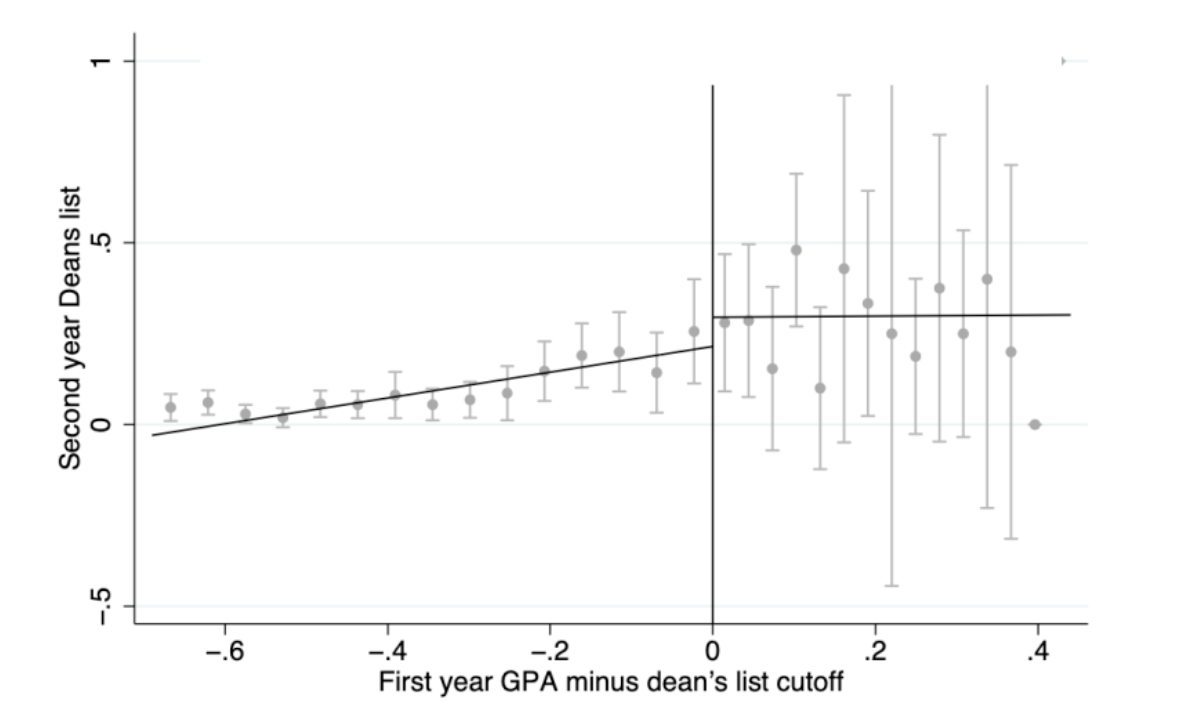} 
\caption{Estimated Impact on the Probability of Making the Dean's List in the Second Year for Low Prior Academic Performance Group}
\label{fig:figure3}
\end{figure}

Students with high school grades below the 50th percentile demonstrate a 10.7\% higher likelihood of making the Dean's List in the second year, controlling for other characteristics (\(p\text{-value} < 0.05\)). This finding suggests a stronger encouragement effect for students with weaker academic backgrounds, aligning with previous studies that highlight the responsiveness of disadvantaged students to educational incentives. Figure~\ref{fig:figure3} further illustrates the discontinuity at the threshold, underscoring the positive impact on low-performing high school students.

\begin{figure}[h!]
\centering
\includegraphics[width=\linewidth]{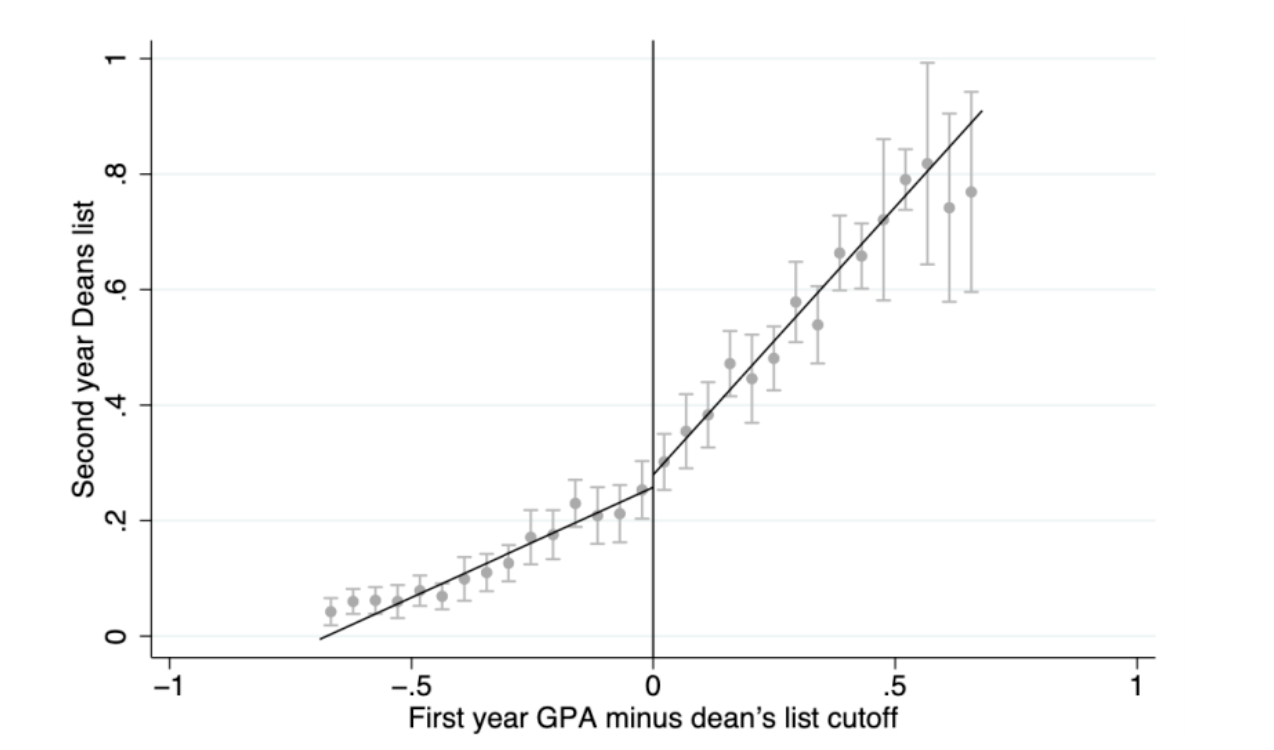} 
\caption{Estimated Impact on the Probability of Making the Dean's List in the Second Year for Native English Speaker}
\label{fig:figure4}
\end{figure}

In contrast, the effect is negligible for students with high school grades above the median. Similarly, native English speakers exhibit a 3.6\% higher likelihood of making the Dean's List in the second year (\(p\text{-value} < 0.10\)), as shown in Figure~\ref{fig:figure4}. This advantage may stem from their proficiency in academic language, which facilitates better comprehension and performance.

\begin{table*}[h!]
\centering
\caption{The Effect of Dean's List on the Probability of Getting Dean's List in the Second Year}
\label{tab:table3}
\def\sym#1{\ifmmode^{#1}\else\(^{#1}\)\fi}
\begin{tabular}{l*{7}{c}}
\hline\hline
\addlinespace
                & \multicolumn{1}{c}{(1)} & \multicolumn{1}{c}{(2)} & \multicolumn{1}{c}{(3)} & \multicolumn{1}{c}{(4)} & \multicolumn{1}{c}{(5)} & \multicolumn{1}{c}{(6)} & \multicolumn{1}{c}{(7)} \\
                & \multicolumn{1}{c}{\makecell{All}} & \multicolumn{1}{c}{\makecell{HS grades \\ below \\ median}} & \multicolumn{1}{c}{\makecell{HS grades \\ above \\ median}} & \multicolumn{1}{c}{Male} & \multicolumn{1}{c}{Female} & \multicolumn{1}{c}{\makecell{Native \\ English \\ speaker}} & \multicolumn{1}{c}{\makecell{Non-native \\ English \\ speaker}} \\ 
\midrule
First year GPA $>$ cutoff & 0.017 & 0.107\sym{**} & 0.013 & -0.003 & 0.032 & 0.036\sym{*} & -0.032 \\
                         & (0.02) & (0.049) & (0.021) & (0.029) & (0.022) & (0.021) & (0.033) \\
Constant                 & 0.23\sym{***} & 0.184\sym{***} & 0.237\sym{***} & 0.231\sym{***} & 0.23\sym{***} & 0.246\sym{***} & 0.183\sym{***} \\
                         & (0.008) & (0.023) & (0.009) & (0.012) & (0.011) & (0.01) & (0.019) \\
Observations             & 10978 & 1709 & 9269 & 4304 & 6674 & 8071 & 2907 \\
\hline\hline
\multicolumn{8}{l}{\footnotesize Notes: Standard errors in parentheses, clustered on GPA.} \\
\multicolumn{8}{l}{\footnotesize \sym{*} \(p<0.10\), \sym{**} \(p<0.05\), \sym{***} \(p<0.01\).} \\
\end{tabular}
\end{table*}

\subsection{Impact on Second-Year GPA}

\begin{figure}[ht]
\centering
\includegraphics[width=\linewidth]{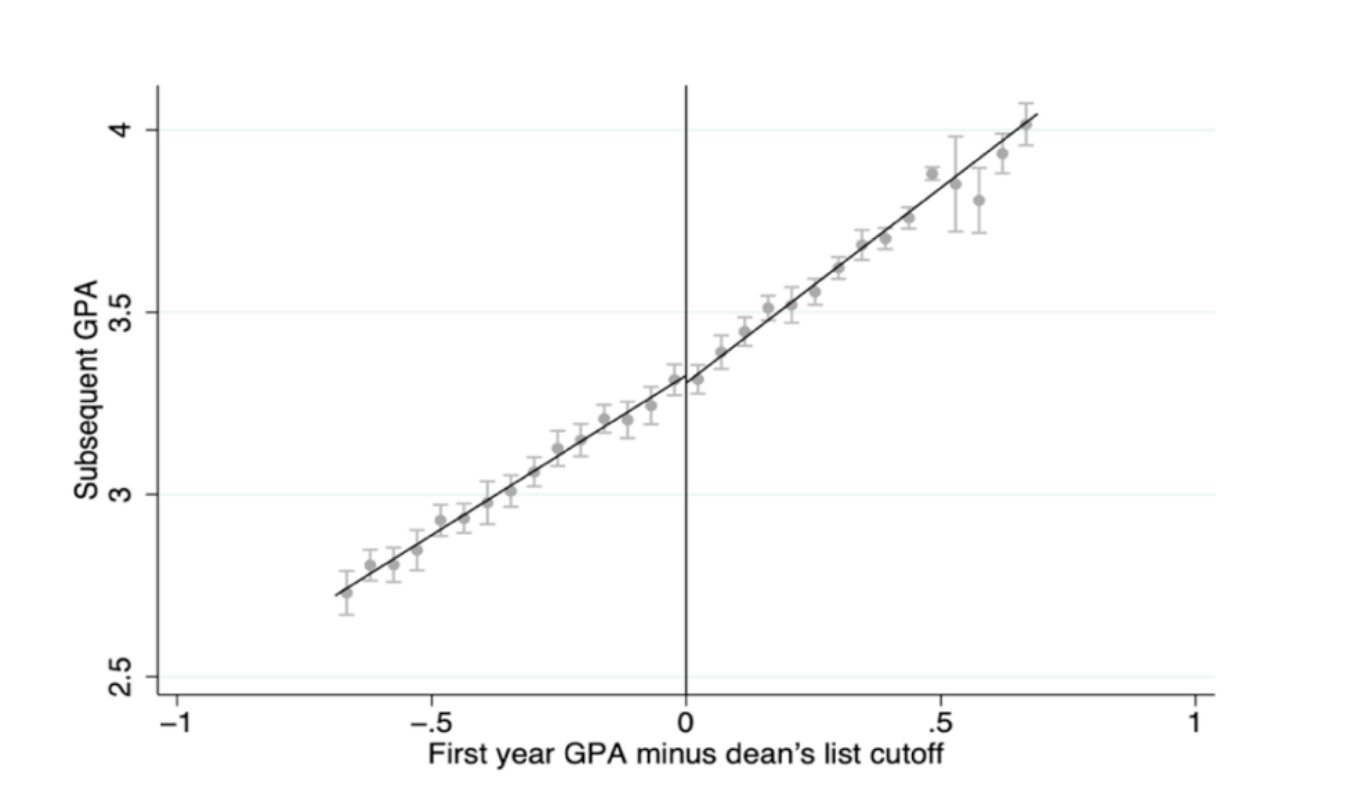} 
\caption{Estimated Effect of Dean’s List on Student Second Year GPA}
\label{fig:figure5}
\end{figure}

Table~\ref{tab:table4} and Figure~\ref{fig:figure5} present the effect of Dean's List placement on second-year GPA. On average, being on the Dean's List results in a 0.018-point decrease in GPA, although this effect is statistically insignificant. Subgroup analyses reveal consistent findings, with no significant variation based on high school performance, gender, or English proficiency.

Figure~\ref{fig:figure6} provides a detailed breakdown by subgroup, confirming the negligible impact across all categories. These results suggest that while Dean's List recognition is a marker of past academic achievement, it does not appear to drive improvements in academic performance in subsequent years.

\begin{table*}[h!]
\centering
\caption{The Effect of Dean's List on Second-Year GPA}
\label{tab:table4}
\def\sym#1{\ifmmode^{#1}\else\(^{#1}\)\fi}
\begin{tabular}{l*{7}{c}}
\hline\hline
\addlinespace
                & \multicolumn{1}{c}{(1)} & \multicolumn{1}{c}{(2)} & \multicolumn{1}{c}{(3)} & \multicolumn{1}{c}{(4)} & \multicolumn{1}{c}{(5)} & \multicolumn{1}{c}{(6)} & \multicolumn{1}{c}{(7)} \\
                & \multicolumn{1}{c}{\makecell{All}} & \multicolumn{1}{c}{\makecell{HS grades \\ below \\ median}} & \multicolumn{1}{c}{\makecell{HS grades \\ above \\ median}} & \multicolumn{1}{c}{Male} & \multicolumn{1}{c}{Female} & \multicolumn{1}{c}{\makecell{Native \\ English \\ speaker}} & \multicolumn{1}{c}{\makecell{Non-native \\ English \\ speaker}} \\ 
\midrule
First year GPA $>$ cutoff & -0.018 & -0.072 & -0.014 & -0.014 & -0.019 & -0.003 & -0.055 \\
                         & (0.02) & (0.069) & (0.019) & (0.031) & (0.021) & (0.022) & (0.038) \\
Constant                 & 3.326\sym{***} & 3.23\sym{***} & 3.334\sym{***} & 3.293\sym{***} & 3.346\sym{***} & 3.336\sym{***} & 3.293\sym{***} \\
                         & (0.013) & (0.03) & (0.013) & (0.018) & (0.016) & (0.014) & (0.031) \\
Observations             & 9979 & 1559 & 8420 & 3927 & 6052 & 7348 & 2631 \\
\hline\hline
\multicolumn{8}{l}{\footnotesize Notes: Standard errors in parentheses, clustered on GPA.} \\
\multicolumn{8}{l}{\footnotesize \sym{*} \(p<0.10\), \sym{**} \(p<0.05\), \sym{***} \(p<0.01\).} \\
\end{tabular}
\end{table*}

\subsection{Effect on Second-Year Credit Load}

The analysis of second-year credit load, summarized in Table~\ref{tab:table5}, shows a negative impact of Dean's List recognition on the number of credits taken. Students who made the Dean's List took an average of 0.029 fewer credits in the following year, with similar trends observed across subgroups. However, none of these effects are statistically significant. This indicates that being on the Dean's List does not motivate students to increase their academic workload in the subsequent year.

\begin{table*}[ht]
\centering
\caption{The Effect of Dean's List on the Total Credit Taken in the Second Year}
\label{tab:table5}
\def\sym#1{\ifmmode^{#1}\else\(^{#1}\)\fi}
\begin{tabular}{l*{7}{c}}
\hline\hline
\addlinespace
                &\multicolumn{1}{c}{(1)}&\multicolumn{1}{c}{(2)}&\multicolumn{1}{c}{(3)}&\multicolumn{1}{c}{(4)}&\multicolumn{1}{c}{(5)}&\multicolumn{1}{c}{(6)}&\multicolumn{1}{c}{(7)}\\
                &\multicolumn{1}{c}{\makecell{All}}&
                \multicolumn{1}{c}{\makecell{HS grades \\ below \\median}}&
                \multicolumn{1}{c}{\makecell{HS grades \\ above \\median}}&
                \multicolumn{1}{c}{Male}&
                \multicolumn{1}{c}{Female}&
                \multicolumn{1}{c}{\makecell{Native \\ English \\ speaker}}&
                \multicolumn{1}{c}{\makecell{Nonnative \\ English \\ speaker}}\\
\midrule
First year GPA $>$ cutoff & -0.029 & -0.018 & -0.018 & -0.028 & -0.030 & -0.043 & -0.006 \\
                          & (0.086) & (0.193) & (0.081) & (0.096) & (0.099) & (0.091) & (0.108) \\
\addlinespace
Constant                  & 3.185\sym{***} & 3.406\sym{***} & 3.153\sym{***} & 3.225\sym{***} & 3.159\sym{***} & 3.159\sym{***} & 3.262\sym{***} \\
                          & (0.061) & (0.111) & (0.058) & (0.074) & (0.064) & (0.059) & (0.079) \\
\midrule
Observations              & 10978 & 1709 & 9269 & 4304 & 6674 & 8071 & 2907 \\
\hline\hline
\multicolumn{8}{l}{\footnotesize Notes: Standard errors in parentheses, clustered on GPA.}\\
\multicolumn{8}{l}{\footnotesize \sym{*} \(p<0.10\), \sym{**} \(p<0.05\), \sym{***} \(p<0.01\).}\\
\end{tabular}
\end{table*}

\begin{table*}[ht]
\centering
\caption{The Effect of Dean’s List on the Dropout Rate}
\label{tab:table6}
\def\sym#1{\ifmmode^{#1}\else\(^{#1}\)\fi}
\begin{tabular}{l*{7}{c}}
\hline\hline
\addlinespace
                &\multicolumn{1}{c}{(1)}&
                 \multicolumn{1}{c}{(2)}&
                 \multicolumn{1}{c}{(3)}&
                 \multicolumn{1}{c}{(4)}&
                 \multicolumn{1}{c}{(5)}&
                 \multicolumn{1}{c}{(6)}&
                 \multicolumn{1}{c}{(7)}\\
                &\multicolumn{1}{c}{\makecell{All}}&
                 \multicolumn{1}{c}{\makecell{HS grades \\ below \\ median}}&
                 \multicolumn{1}{c}{\makecell{HS grades \\ above \\ median}}&
                 \multicolumn{1}{c}{Male}&
                 \multicolumn{1}{c}{Female}&
                 \multicolumn{1}{c}{\makecell{Native \\ English \\ speaker}}&
                 \multicolumn{1}{c}{\makecell{Nonnative \\ English \\ speaker}}\\
\midrule
First year GPA $>$ cutoff & 0.000 & -0.012 & 0.000 & 0.009 & -0.008 & 0.007 & -0.020 \\
                          & (0.014) & (0.037) & (0.014) & (0.017) & (0.018) & (0.015) & (0.023) \\
\addlinespace
Constant                  & 0.094\sym{***} & 0.092\sym{***} & 0.094\sym{***} & 0.095\sym{***} & 0.094\sym{***} & 0.091\sym{***} & 0.106\sym{***} \\
                          & (0.008) & (0.019) & (0.008) & (0.011) & (0.010) & (0.009) & (0.015) \\
\midrule
Observations              & 10978 & 1709 & 9269 & 4304 & 6674 & 8071 & 2907 \\
\hline\hline
\multicolumn{8}{l}{\footnotesize Notes: Standard errors in parentheses, clustered on GPA.}\\
\multicolumn{8}{l}{\footnotesize \sym{*} \(p<0.10\), \sym{**} \(p<0.05\), \sym{***} \(p<0.01\).}\\
\end{tabular}
\end{table*}

\subsection{Impact on Retention and Dropout Rates}

\begin{figure}[!ht]
\centering
\includegraphics[width=0.85\linewidth]{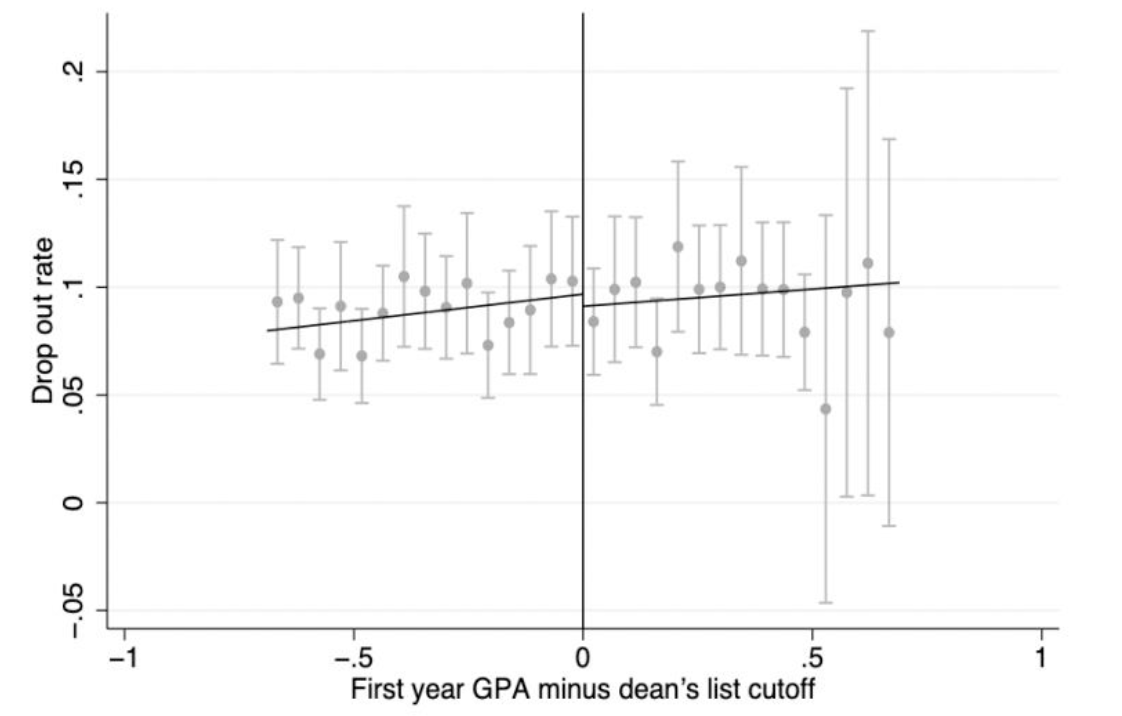} 
\caption{Estimated Effect of Being on the Dean's List on Decision to Drop Out.}
\label{fig:figure7}
\end{figure}

The study also investigates the relationship between Dean's List recognition and student retention, focusing on dropout rates. Table~\ref{tab:table6} and Figure~\ref{fig:figure7} show no significant effect of Dean's List placement on the likelihood of dropping out. While the overall point estimate is negative, suggesting a potential reduction in dropout rates, it is not statistically significant for any subgroup. These findings imply that Dean's List recognition does not play a decisive role in student retention.

\subsection{Probability of Graduating Within Four Years}

\begin{figure}[!ht]
\centering
\includegraphics[width=0.85\linewidth]{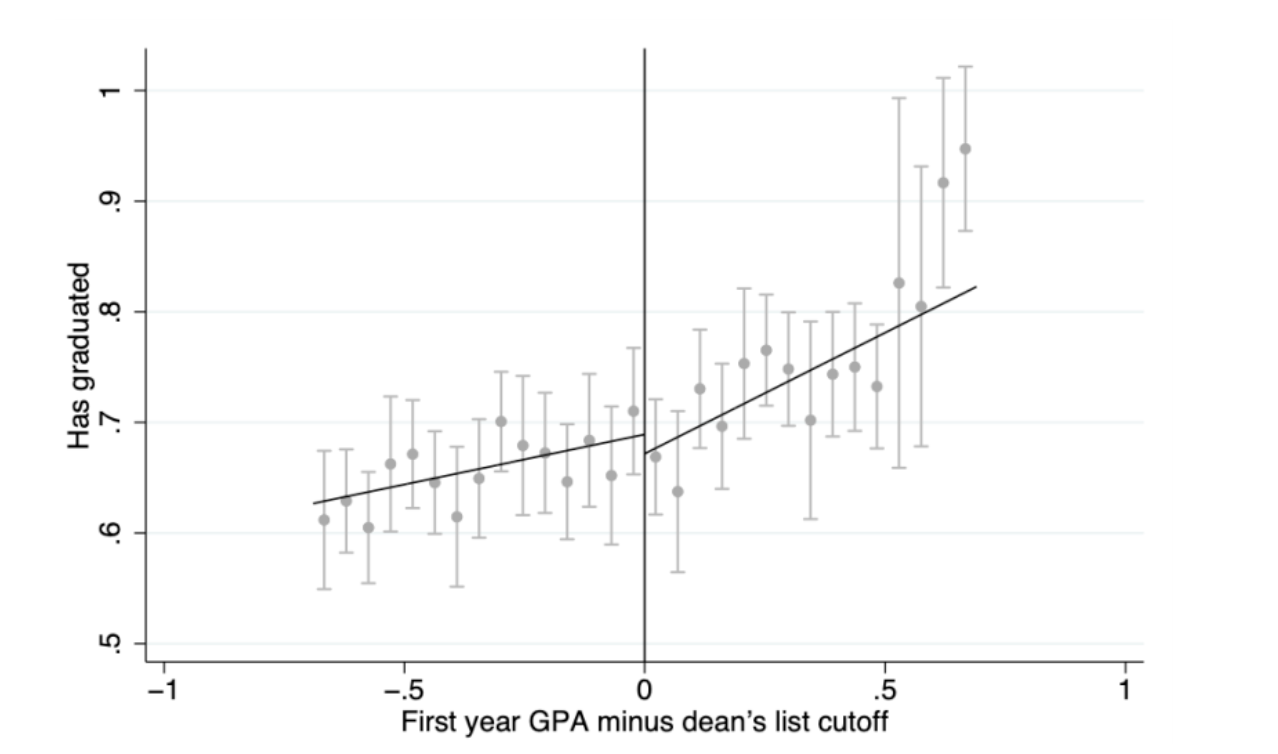} 
\caption{Estimated Effect of Being on the Dean's List on Probability of Graduating within Four Years.}
\label{fig:figure8}
\end{figure}

\begin{figure*}[!ht]
\centering
\subfloat[High School Grade $<$ Median]{%
    \includegraphics[width=0.45\linewidth]{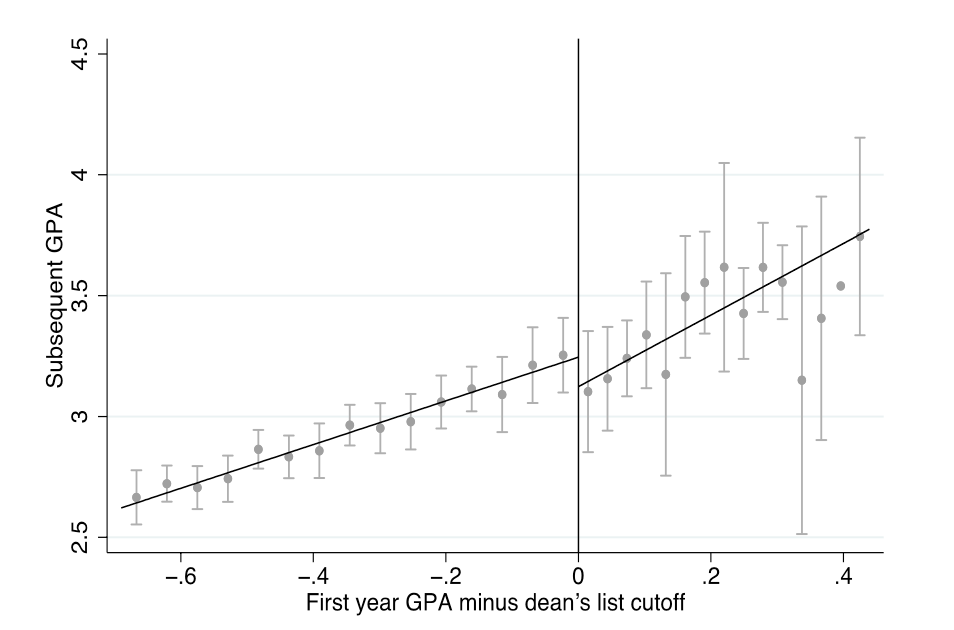}
    \label{fig:6a}
}
\hfill
\subfloat[High School Grade $>$ Median]{%
    \includegraphics[width=0.45\linewidth]{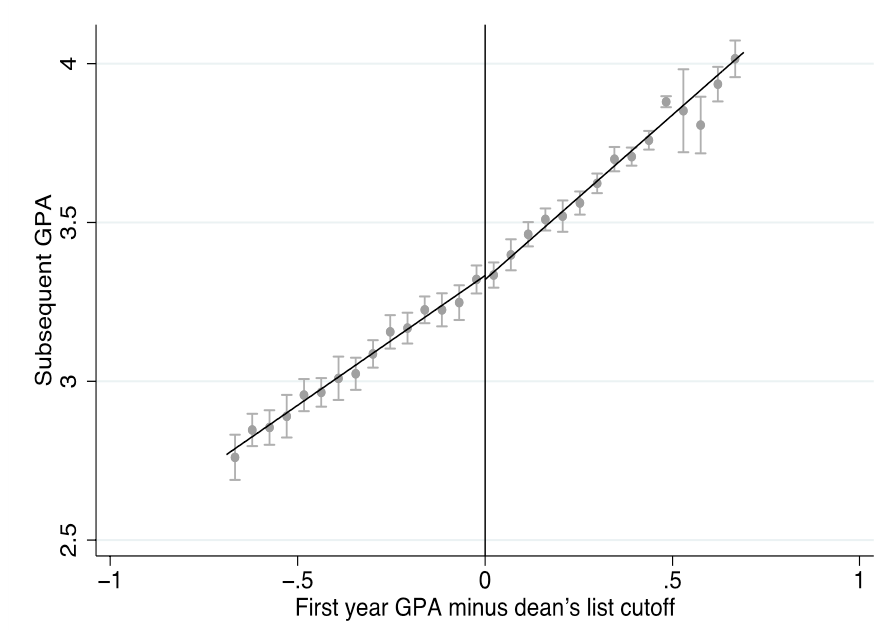}
    \label{fig:6b}
}
\\ 
\subfloat[Male]{%
    \includegraphics[width=0.45\linewidth]{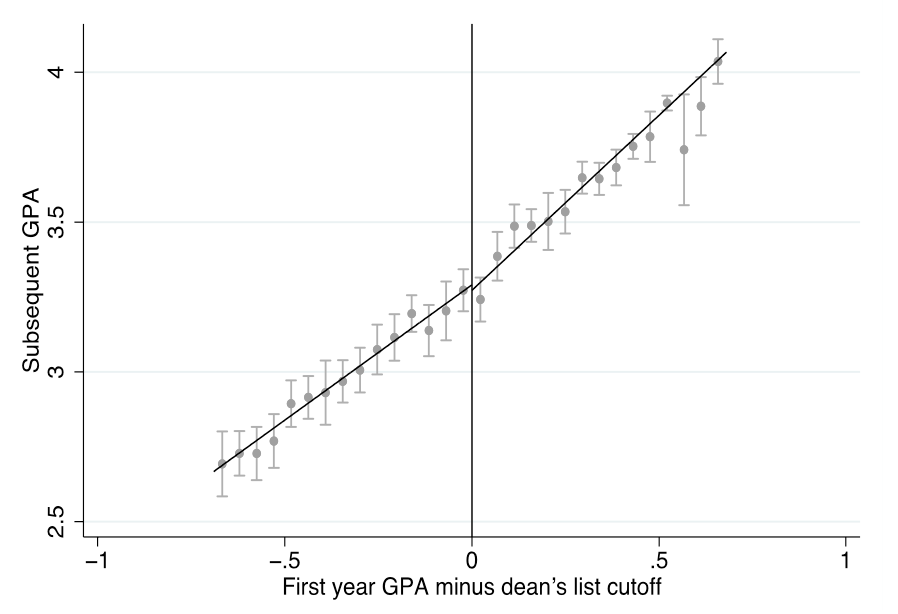}
    \label{fig:6c}
}
\hfill
\subfloat[Female]{%
    \includegraphics[width=0.45\linewidth]{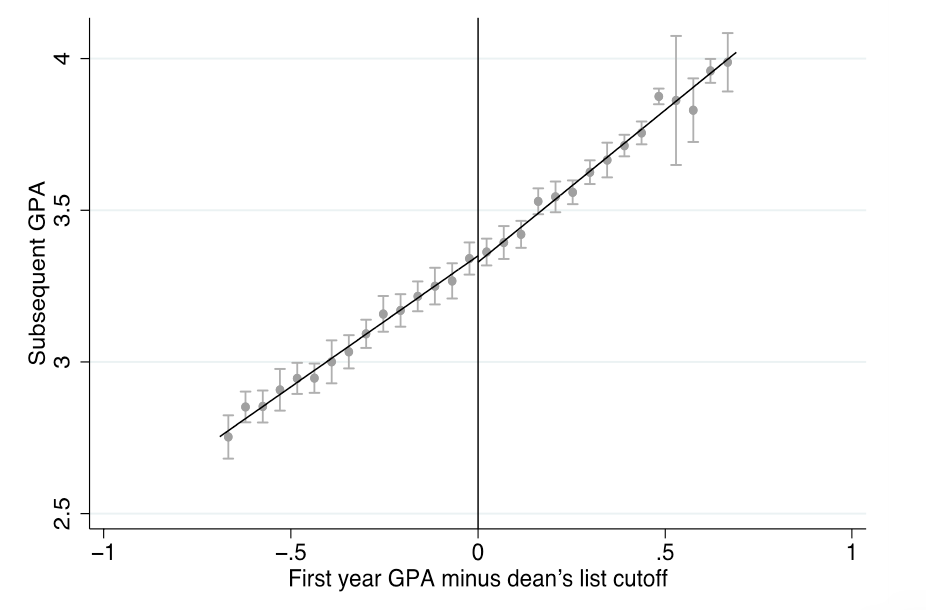}
    \label{fig:6d}
}
\\ 
\subfloat[Native Speaker]{%
    \includegraphics[width=0.45\linewidth]{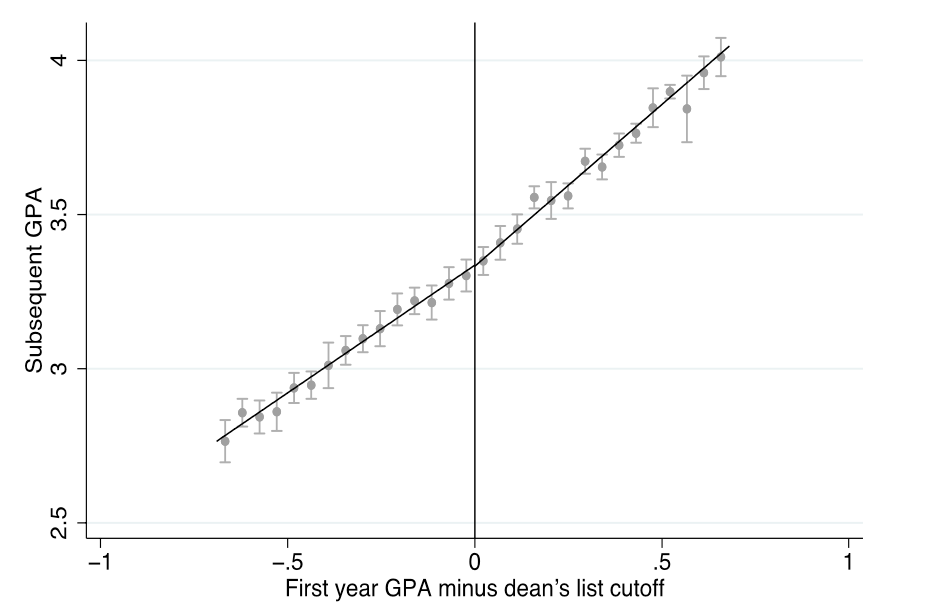}
    \label{fig:6e}
}
\hfill
\subfloat[Non-native Speaker]{%
    \includegraphics[width=0.45\linewidth]{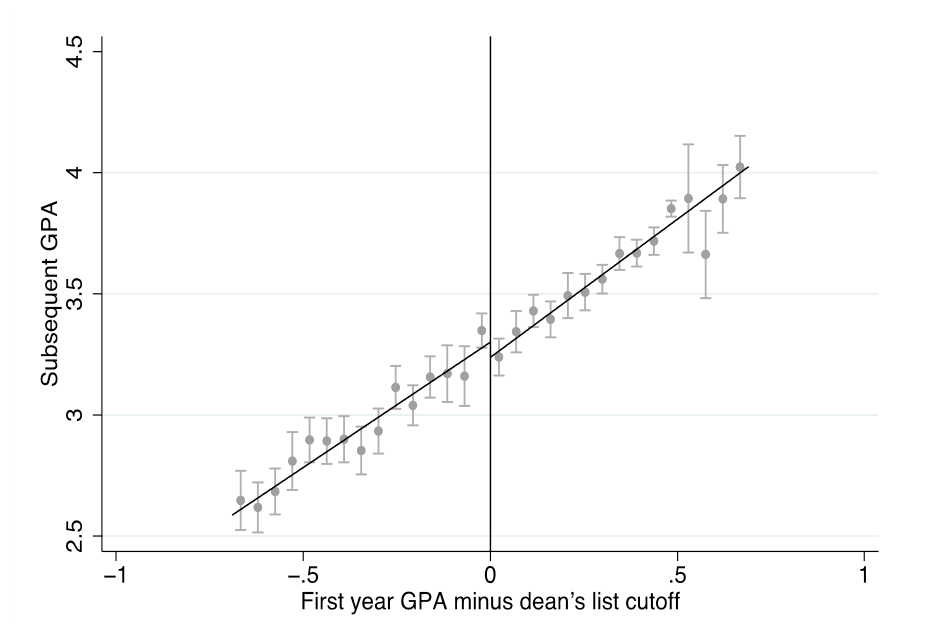}
    \label{fig:6f}
}
\caption{Estimated Effect of Being Named to the Dean's List on Second-Year GPA for Subgroups of Students.}
\label{fig:figure6}
\end{figure*}

To assess the long-term impact of Dean's List recognition, the study examines its effect on the probability of graduating within four years. Table~\ref{tab:table7} and Figure~\ref{fig:figure8} reveal a slight negative effect, with an average reduction of 2.2\% in the likelihood of graduating within four years. However, this result is not statistically significant, and no meaningful variation is observed across subgroups.

\subsection{Summary of Findings}

The results of this study suggest that while Dean's List recognition has a positive impact on the likelihood of making the list again for certain subgroups—particularly low-performing high school students and native English speakers—it does not translate into significant improvements in second-year GPA, credit load, retention, or graduation rates. These findings emphasize the need for institutions to explore alternative strategies for fostering sustained academic success beyond symbolic recognition.

\begin{table*}[!ht]
\centering
\caption{The Effect of Dean's List on the Probability of Graduating within Four Years}
\label{tab:table7}
\def\sym#1{\ifmmode^{#1}\else\(^{#1}\)\fi}
\begin{tabular}{l*{7}{c}}
\hline\hline
\addlinespace
                &\multicolumn{1}{c}{(1)}&
                 \multicolumn{1}{c}{(2)}&
                 \multicolumn{1}{c}{(3)}&
                 \multicolumn{1}{c}{(4)}&
                 \multicolumn{1}{c}{(5)}&
                 \multicolumn{1}{c}{(6)}&
                 \multicolumn{1}{c}{(7)}\\
                &\multicolumn{1}{c}{\makecell{All}}&
                 \multicolumn{1}{c}{\makecell{HS grades \\ below \\ median}}&
                 \multicolumn{1}{c}{\makecell{HS grades \\ above \\ median}}&
                 \multicolumn{1}{c}{Male}&
                 \multicolumn{1}{c}{Female}&
                 \multicolumn{1}{c}{\makecell{Native \\ English \\ speaker}}&
                 \multicolumn{1}{c}{\makecell{Nonnative \\ English \\ speaker}}\\
\midrule
First year GPA $>$ cutoff & -0.022 & -0.015 & -0.029 & -0.032 & -0.009 & -0.024 & -0.016 \\
                          & (0.026) & (0.108) & (0.024) & (0.040) & (0.030) & (0.027) & (0.052) \\
\addlinespace
Constant                  & 0.692\sym{***} & 0.624\sym{***} & 0.704\sym{***} & 0.649\sym{***} & 0.720\sym{***} & 0.694\sym{***} & 0.686\sym{***} \\
                          & (0.013) & (0.033) & (0.014) & (0.023) & (0.016) & (0.014) & (0.035) \\
\midrule
Observations              & 7167 & 1202 & 5965 & 2828 & 4339 & 5404 & 1763 \\
\hline\hline
\multicolumn{8}{l}{\footnotesize Notes: Standard errors in parentheses, clustered on GPA.}\\
\multicolumn{8}{l}{\footnotesize \sym{*} \(p<0.10\), \sym{**} \(p<0.05\), \sym{***} \(p<0.01\).}\\
\end{tabular}
\end{table*}

\newpage

\section{Discussion}

The analysis of various academic outcomes, including second-year GPA, credit load, and the likelihood of dropping out, reveals nuanced effects of being on the Dean's List. The findings indicate that students with poor high school academic performance are significantly more likely to make the Dean's List again in the second year. This aligns with prior research suggesting that students from disadvantaged academic backgrounds may be more responsive to positive educational incentives. A plausible explanation for this finding is that these students experience a heightened sense of accomplishment and recognition from being on the Dean's List, which motivates them to sustain or improve their academic performance.

Similarly, the analysis shows a moderately positive impact on native English speakers’ likelihood of being recognized on the Dean's List in the subsequent year. This advantage may stem from their familiarity with academic language, enabling them to excel in coursework and maintain higher grades. However, the study finds no significant causal effect of being on the Dean's List on subsequent GPA, student retention, or the likelihood of graduating within four years, regardless of students’ characteristics. These findings suggest that being on the Dean's List serves primarily as recognition of prior achievement rather than a driver of future academic success.

The lack of a sustained academic impact highlights several potential limitations. First, students who make the Dean's List may already be high achievers, with their future performance more influenced by intrinsic factors like motivation, study habits, and critical thinking skills rather than external recognition. Second, the Dean's List may not provide sufficient tangible benefits to serve as a strong motivator for continued excellence. For instance, students motivated by external recognition or resume-building may not find the designation compelling enough to drive significant changes in their academic effort.

Additionally, insights from anomaly detection methodologies could be leveraged to identify students whose academic trajectories deviate significantly from expected patterns. Applying anomaly detection techniques to the context of Dean's List recognition could offer a deeper understanding of outlier behaviors, shedding light on students who respond to incentives in unexpected ways\cite{hu2025efficient}. This, in turn, could help inform more targeted and effective interventions to support their academic success.

Additionally, this study's reliance on data limited to students' first and second years constrains its ability to fully assess the long-term effects of being on the Dean's List. Graduation rates and other long-term outcomes are influenced by a range of factors, including third- and fourth-year academic performance, personal circumstances, and external pressures like employment opportunities and financial challenges. Without access to data beyond the second year, the findings regarding the long-term impact of the Dean's List on student outcomes remain inconclusive.

\section{Conclusion}

This study employed regression discontinuity analysis to estimate the causal impact of being on the Dean's List on students’ future academic performance. The findings reveal that being on the Dean's List has a positive effect on the probability of making the list again in the second year, particularly for students with poor high school academic performance and native English speakers. However, no significant effects were observed on subsequent GPA, credit load, dropout rates, or the likelihood of graduating within four years.

While the Dean's List may serve as an important symbol of academic achievement, its role in fostering future academic success appears limited. The variation in effects across subgroups underscores the need for a more tailored approach to academic recognition. Students from disadvantaged backgrounds or those struggling academically may benefit from additional support, while high-achieving students may require other types of incentives to sustain their performance.

These findings have practical implications for higher education institutions. While symbolic recognition like the Dean's List is valuable, institutions should explore more impactful ways to incentivize academic excellence. This could include providing additional resources, such as access to research opportunities, mentorship programs, or leadership roles, to foster sustained academic growth. 

In conclusion, the study demonstrates that while being on the Dean's List can encourage specific groups of students, it does not have a significant causal impact on academic success overall. Further research should explore other factors, such as motivation, study habits, and institutional support systems, that contribute to long-term academic achievement. Institutions must also consider innovative strategies to recognize and reward academic excellence beyond symbolic designations, ensuring a meaningful and lasting impact on students' educational journeys.

\bibliographystyle{IEEEtran}
\bibliography{references}

\end{document}